\documentclass[12pt,preprint]{aastex}
\usepackage{epsf}
\usepackage{graphicx}
\usepackage{lscape}
\usepackage{natbib}
\usepackage{array}

\tabletypesize{\scriptsize}
\newcommand{\etal}{{et al.}\ }

\def\gappeq{\mathrel{ \rlap{\raise.5ex\hbox{$>$}}
                      {\lower.5ex\hbox{$\sim$}}  } }

\defcitealias{c99}{C99}
\defcitealias{mmm02}{MMM02}
\defcitealias{dll98}{DLL98}

\begin{document}
\shorttitle{Exozodiacal Structure Formed by Earth-Mass Planets}
\shortauthors{Stark \etal}

\title{The Detectability of Exo-Earths and Super-Earths Via Resonant Signatures in Exozodiacal Clouds}

\author{Christopher C. Stark\altaffilmark{1} and Marc J. Kuchner\altaffilmark{2}}

\altaffiltext{1}{Department of Physics, University of Maryland, Box 197, 082 Regents Drive,
College Park, MD 20742-4111, USA;
starkc@umd.edu}
\altaffiltext{2}{NASA Goddard Space Flight Center, Exoplanets and Stellar
Astrophysics Laboratory, Code 667, Greenbelt, MD 20771}

\begin{abstract}

Directly imaging extrasolar terrestrial planets necessarily means contending with the astrophysical noise of exozodiacal dust and the resonant structures created by these planets in exozodiacal clouds.  Using a custom tailored hybrid symplectic integrator we have constructed 120 models of resonant structures created by exo-Earths and super-Earths on circular orbits interacting with collisionless steady-state dust clouds around a Sun-like star.  Our models include enough particles to overcome the limitations of previous simulations that were often dominated by a handful of long-lived particles, allowing us to quantitatively study the contrast of the resulting ring structures.  We found that in the case of a planet on a circular orbit, for a given star and dust source distribution, the morphology and contrast of the resonant structures depend on only two parameters: planet mass and $\sqrt{a_{\rm p}}/\beta$, where $a_{\rm p}$ is the planet's semi-major axis and $\beta$ is the ratio of radiation pressure force to gravitational force on a grain.  We constructed multiple-grain-size models of 25,000 particles each and showed that in a collisionless cloud, a Dohnanyi crushing law yields a resonant ring whose optical depth is dominated by the largest grains in the distribution, not the smallest.  We used these models to estimate the mass of the lowest-mass planet that can be detected through observations of a resonant ring for a variety of assumptions about the dust cloud and the planet's orbit.  Our simulations suggest that planets with mass as small as a few times Mar's mass may produce detectable signatures in debris disks at $a_{\rm p} \gtrsim 10$ AU.

\end{abstract}


\section{Introduction}

A number of proposed experiments like the Terrestrial Planet Finder (TPF) aim to directly image the scattered and emitted light from extrasolar planets \citep{lt06}.  These experiments will also excel at detecting exozodiacal dust, circumstellar dust analogous to zodiacal dust in our solar system \citep[e.g.][]{a07,b07}.  Zodiacal dust in the solar system consists of $\sim1-100 \; \mu$m dust grains released through asteroidal collisions and the outgassing of comets \citep[e.g.][]{sbw89}.  This dust forms the zodiacal cloud, extending from the solar corona \citep[e.g.][]{mkk00} to beyond Jupiter \citep[e.g.][]{k99}.

Our zodiacal cloud exhibits several structures interpreted as dynamical signatures of planets \citep{d85, d94, r95}.  Several dusty disks around nearby main-sequence stars show similar structures \citep[e.g.][]{g98, w02, kgc05}.  This trend suggests that exozodiacal clouds may be full of rings, clumps and other asymmetries caused by planets and other phenomena.

This situation raises some important questions.  Will the structures in exozodiacal clouds be harmful astrophysical noise for direct imaging of extrasolar planets \citep{b96, bwl99}?  Or can the dynamical signatures of planets in these clouds help us find otherwise undetectable planets \citep[e.g.][]{kh03}?

Several studies have examined the geometry of resonant signatures of planets in debris disks \citep[e.g.][]{kh03, r08}.  However, most simulations cannot quantitatively study the contrast in these structures: how bright they are relative to the background cloud.  We need to model the contrast of the structures in exozodiacal clouds to understand their roles as astrophysical noise and as signposts of hidden planets.  However, accurately simulating the contrast of these structures demands computational resources that have only recently become available \citep[e.g.][]{dm05}.

In this paper we examine the contrast of resonant structures induced by planets in steady-state exozodiacal clouds and the detectability of these structures via direct imaging.  We simulate high-fidelity images of collisionless exozodiacal clouds containing a terrestrial-mass planet---an exo-Earth or super-Earth.  By using roughly an order of magnitude more particles than most previous simulations, we overcome the Poisson noise associated with constructing histograms of the column density and populating the external mean motion resonances (MMRs) of planets.  We use our simulations to estimate the minimum planet mass that can be indirectly detected via observations of these structures as a function of the planet semi-major axis and dominant grain size under the assumption of circular planet orbits.  Our models apply to exozodiacal clouds less than a few hundred times the optical depth of the solar zodiacal cloud, clouds for which the collision time is shorter than the Poynting-Robertson (PR) time for typical grains.

Section 2 of this paper describes our numerical techniques.  We present a synthetic catalog of resonant debris disk structures in Section 3.  We describe our multiple-particle-size cloud models and discuss their detectability in Section 4.  In Section 5, we discuss the limitations of our simulations; we summarize our conclusions in Section 6.

\section{Numerical Method}

Dust grains in the inner solar system are primarily released from parent bodies via collisions or outgassing.  Radiation pressure ejects the smallest particles from the solar system in a dynamical time while the larger particles slowly spiral inward due to PR drag \citep{r37, bls79}.  During their spiral toward the Sun, particles may become temporarily trapped in the MMRs of planets, extending their lifetimes by a factor of a few to ten \citep{jz89}.  This trapping locally enhances the particle density, creating structures within the zodiacal cloud, which have been described as circumsolar rings, bands, and clumps \citep[e.g.][]{k98}.

To model these types of structures in exozodiacal clouds we numerically integrated the equation of motion of dust particles.  The equation of motion for a perfectly absorbing particle orbiting a star of mass $m_{\rm \star}$ is given to first order in $v/c$ by \citet{r37}:

\begin{equation}
  \frac{d^2{\bf r}}{dt^2} = -\frac{Gm_{\rm \star}}{r^2}(1-\beta)\hat{\bf r}-\frac{(1+{\rm sw})\beta}{c}\frac{Gm_{\rm \star}}{r^2}[{\dot{r}}{\hat{\bf r}} + {\bf v}],
\end{equation}

\noindent where $\textbf{r}$ and $\textbf{v}$ are the heliocentric position and velocity of the particle and ${\rm sw}$ is the ratio of solar wind drag to PR drag.  We assume a value for ${\rm sw}$ of 0.35 \citep{g94}.  For perfectly absorbing spherical particles in the vicinity of the Sun, $\beta \approx 0.57/\rho s$, where $\rho$ is the mass density of the particle in ${\rm g \: cm^{-3}}$ and $s$ is the radius in $\mu {\rm m}$.

\subsection{A Customized Hybrid Symplectic Integrator \label{hybridsymp}}

We implemented a customized hybrid symplectic integrator to perform our numerical integrations.  \citet{c99}, hereby referred to as \citetalias{c99}, introduced hybrid symplectic integration as a method for dealing with close encounters in an efficient n-body code.  Symplectic integrators rely on splitting the Hamiltonian into two easily integrable portions---a dominant term, $H_{\rm D}$, and a smaller perturbative term, $H_{\rm P}$.  However, in the n-body problem, $H_{\rm P}$ may exceed $H_{\rm D}$ during close encounters.  Hybrid symplectic integrators overcome this problem by effectively switching from a symplectic integrator to an alternate integrator (e.g. Bulirsch-Stoer).  

The hybrid method reduces the perturbative term of the Hamiltonian, $H_{\rm P}$, by a factor $K(r_{ij})$, where $r_{ij}$ is the distance between the two bodies in question, to ensure that the perturbative term remains relatively small.  The integrator includes the remaining portion of the perturbative term, $\sum_{i,j}H_{{\rm P},ij} [1-K(r_{ij})]$, in the dominant term which is then integrated using a method of choice.  The ``changeover function," $K(r_{ij})$, is a smooth function that varies from 0 for $r_{ij} \lesssim r_{\rm crit}$ to unity for $r_{ij} \gtrsim r_{\rm crit}$.

Using a hybrid integrator requires choosing a changeover function and a value for $r_{\rm crit}$.  We use the same changeover function as \citetalias{c99}.  We assign a different value of $r_{{\rm crit,} i}$ to each body, calculated as the larger of $3R_{{\rm H},i}$ and $\tau v_{i}$, where $R_{{\rm H},i}$ and $v_{i}$ are the Hill radius and velocity of the $i^{\rm th}$ body, respectively, and $\tau$ is the time step of the integrator.  We then calculate the critical distance for a pair of bodies as $r_{{\rm crit},ij}$ = $r_{{\rm crit,} i}$ + $r_{{\rm crit,} j}$.

Our integrator also incorporates the effects of radiation pressure, PR drag, and solar wind drag.  We implement radiation pressure as a correction to the effective stellar mass (cf. Eq. 1) and treat the drag effects as an additional term in $H_{\rm P}$, in much the same way as \citet{mmm02}, hereby referred to as \citetalias{mmm02}.  We also use democratic heliocentric (DH) coordinates, composed of the barycentric momenta and heliocentric positions, because of their relative ease of implementation.  This choice introduces an additional perturbative term to the Hamiltonian due to the motion of the star with respect to the barycenter \citep{dll98}.

\subsection{Comparison of Integrator with Previous Results}

We checked our integrator using a variety of standard tests.  We checked the energy and Jacobi constant conservation with the drag terms turned off and examined the evolution of dust particles' orbital elements under our implementation of drag effects.  We also compared our hybrid integrator to a Bulirsch-Stoer integrator by examining the path of an individual test particle during a close encounter and by examining the statistics of a cloud of particles in a collisionless disk containing a planet.

We tested energy conservation in our integration code by integrating the orbits of the four outer planets and the Sun for $3\times10^5$ years using a time step of 0.15 years.  The energy error was bounded with a mean value of $\Delta{\rm E/E} \approx 3 \times 10^{-9}$.

\begin{figure}
\begin{center}
\includegraphics[width=6.5in]{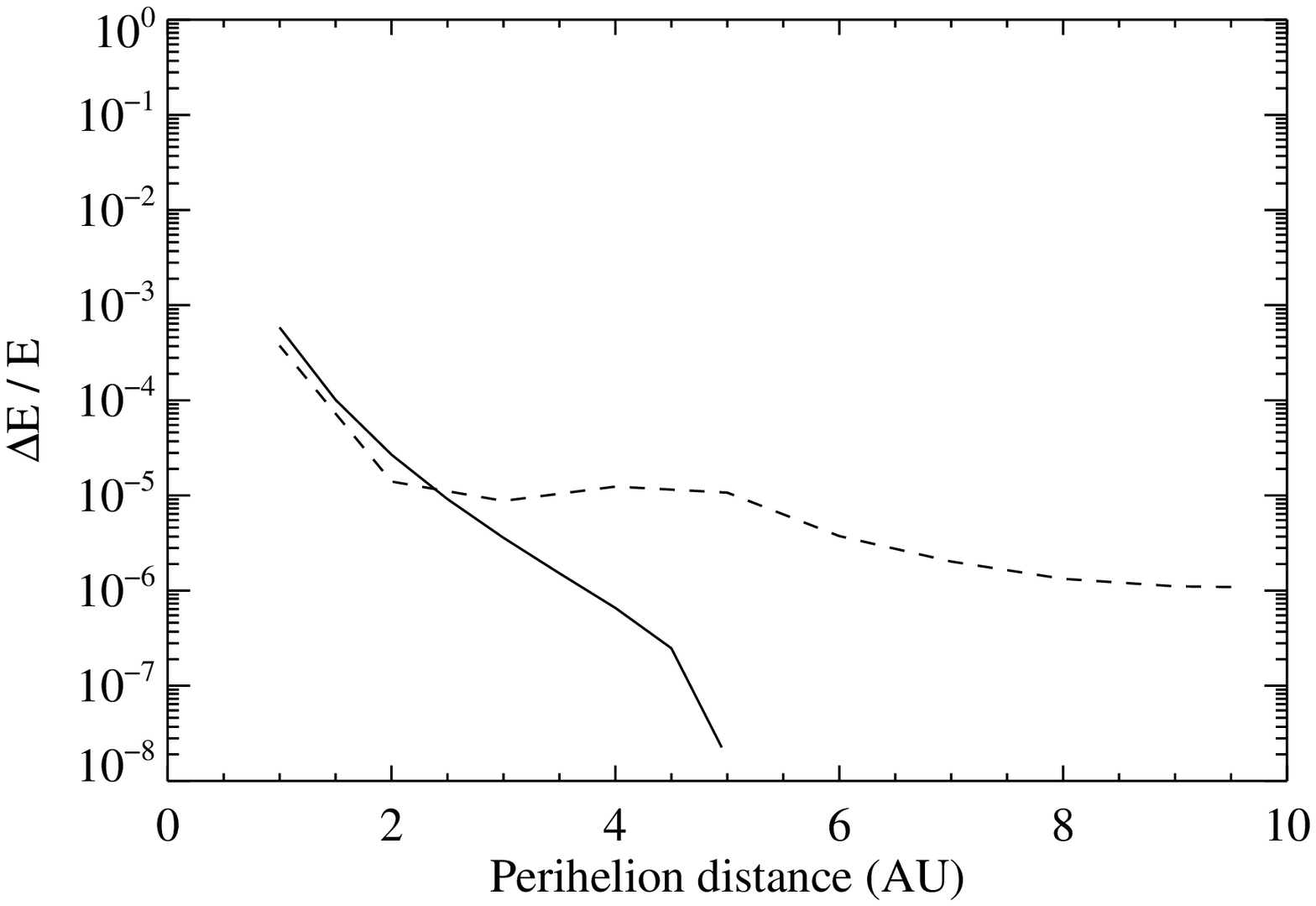}
\caption{Maximum fractional error in energy during a 3,000-year integration as a function of perihelion distance for two scenarios: a two-body system of the Sun \& Jupiter (solid line) and a three-body system of the Sun, Jupiter \& Saturn (dashed line).  For the two-body system, Jupiter's perihelion distance is plotted.  For the three-body system, Saturn's perihelion distance was altered while Jupiter's remained fixed.  The inclinations and eccentricities of both planets remained fixed.  cf. \citetalias{dll98} Fig. 3. \label{energyvsperi}}
\end{center}
\end{figure}

\citet{dll98}, which we will refer to as \citetalias{dll98}, tested the relative conservation of energy in their symplectic integrator as a function of planet perihelion distance.  We replicated their tests using our code.  Figure \ref{energyvsperi} shows the relative energy error in an integration of the orbit of Jupiter for $3\times10^5$ years using a time step of 0.15 years.  We initially placed Jupiter at aphelion.  Figure \ref{energyvsperi} also shows the results of integrating the orbits of Jupiter and Saturn under the same conditions.  With the DH method, the perturbative solar term increases as the perihelion distance of the planet decreases, causing the fractional energy error to increase similarly.  The fractional energy errors shown in Figure \ref{energyvsperi} agree with those obtained by \citetalias{dll98}.

We checked the conservation of the Jacobi constant by integrating particles in the Sun-Neptune system.  We found results consistent with those of \citetalias{mmm02}.  Particles that did not undergo close encounters conserved the Jacobi constant at the level of $\sim 10^{-8}$ to $10^{-7}$.

\begin{figure}
\begin{center}
\includegraphics[width=6.5in]{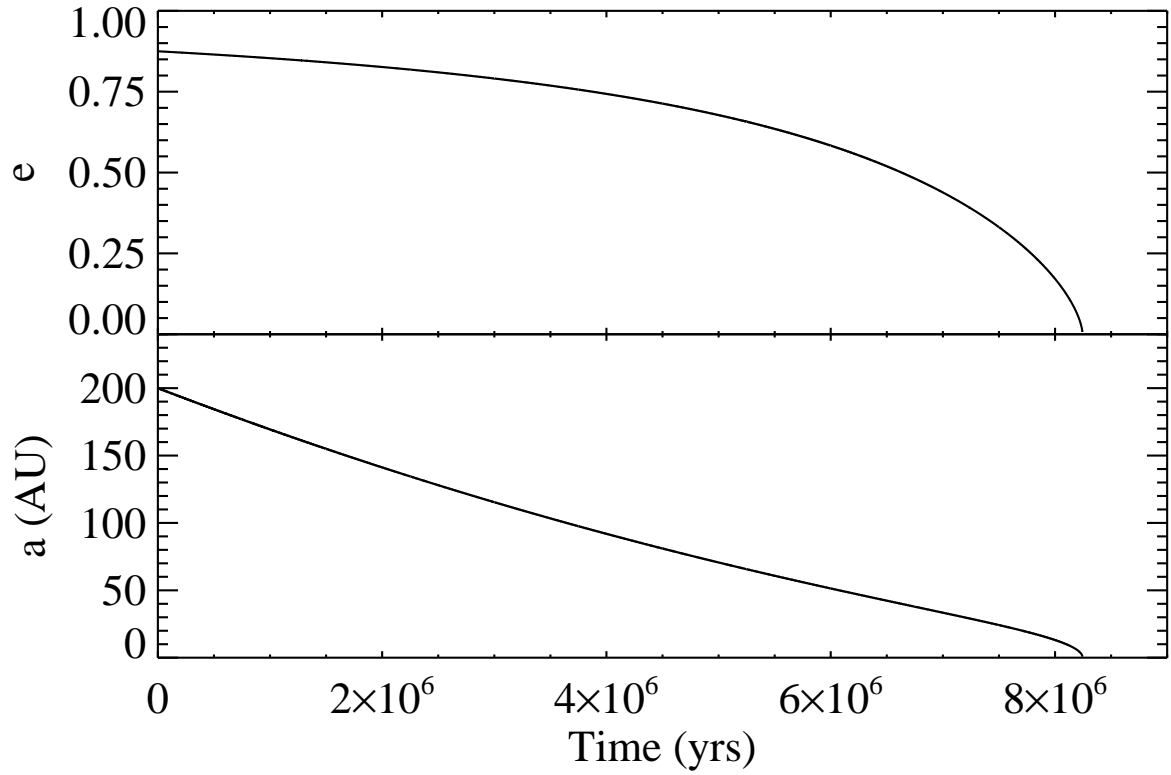}
\caption{\textit{Top:} eccentricity as a function of time for a dust particle with $\beta = 0.2$ and ${\rm sw} = 0.35$.  \textit{Bottom:} semi-major axis as a function of time.   Our results match those of \citetalias{mmm02} (cf. \citetalias{mmm02}, Fig. 1) and agree with the analytical solution \citep{ww50}. \label{aevst}}
\end{center}
\end{figure}

To test our implementation of PR drag, we replicated a test performed by \citetalias{mmm02}.  We integrated the orbit of a particle with $\beta=0.2$ and ${\rm sw}=0.35$ in the presence of the Sun.  Figure \ref{aevst} shows the semi-major axis and eccentricity as functions of time.  These results match the results of \citetalias{mmm02} and agree with the analytic solution \citep{ww50}.

\begin{figure}
\begin{center}
\includegraphics[width=6.5in]{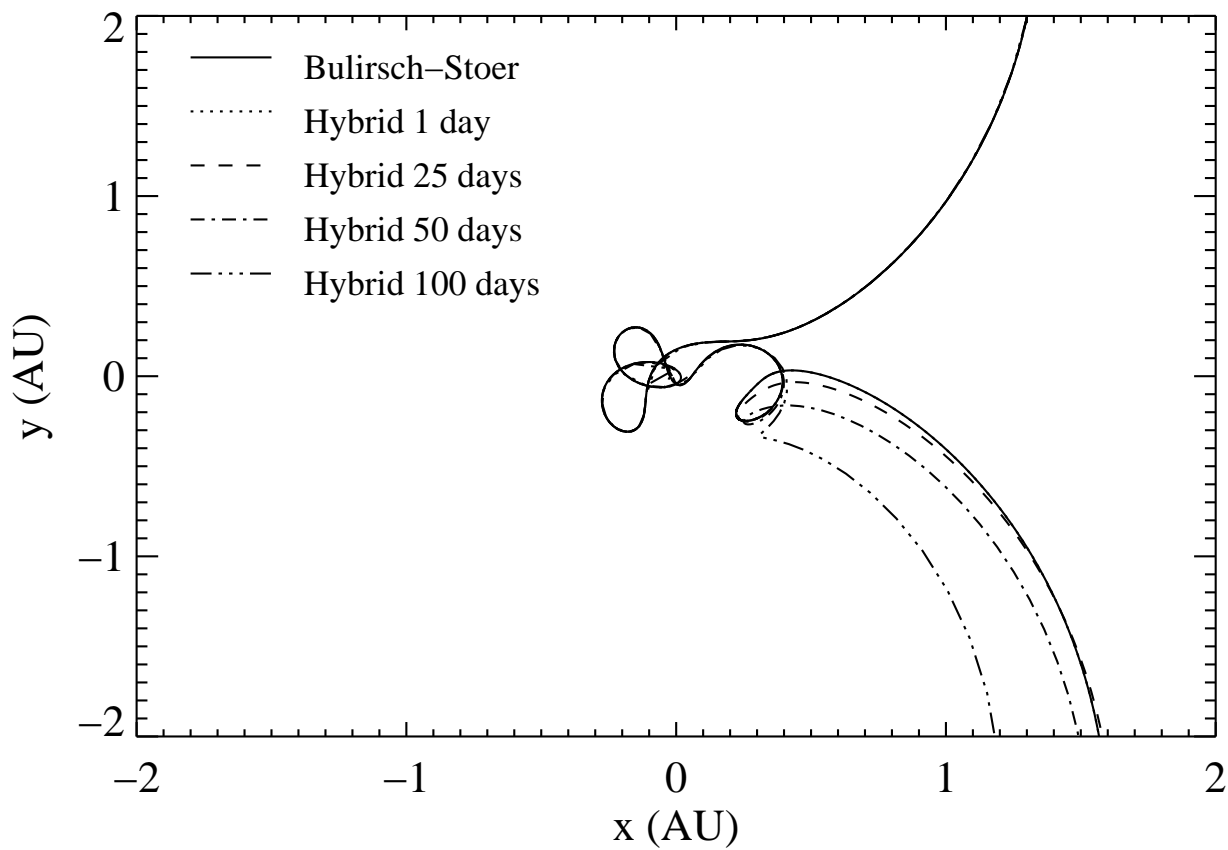}
\caption{The integrated trajectory of comet P/Oterma during a close encounter with Jupiter as viewed in the frame centered on and rotating with Jupiter with the Sun on the negative $x$-axis.  Shown are the results of a Bulirsch-Stoer integrator and our hybrid symplectic integrator for four values of integration time step.  The hybrid symplectic results overlap the Bulirsch-Stoer results for a timestep of 1 day (cf. \citetalias{c99}, Fig. 4). \label{comettrajectory}}
\end{center}
\end{figure}

We tested the performance of our hybrid scheme by integrating the orbit of comet P/Oterma in a close encounter with Jupiter, which has been done previously by \citet{mv96} and \citetalias{c99}.  The initial conditions for both bodies can be found in Table 3 of \citet{mv96}.  Figure \ref{comettrajectory} shows the path of comet P/Oterma for several values of integration time step $\tau$ as seen in the frame co-rotating with Jupiter, which is located at the origin.  These results are similar to those obtained by \citetalias{c99}.  Our code shows a minor improvement over the other codes, most noticeable in the $\tau=100$ days case, that is likely only due to differences in the calculation of the changeover distance.  \citetalias{c99} explicitly sets $r_{\rm crit}=3R_{\rm H}$ for this test; we used our prescription for $r_{\rm crit}$ as described in Section \ref{hybridsymp}.

\subsection{Test Simulations of a Steady-State Exozodiacal Cloud}

We directly compared simulations of resonant structures made with our hybrid integrator to simulations made with a Bulirsch-Stoer integrator.  During the integrations we recorded the coordinates of each particle in a 2-D histogram at regular intervals.   This histogram models the surface density distribution in a steady-state cloud.  Since we only modeled planets on circular orbits, we simply recorded the coordinates in the frame co-rotating with the planet.  This technique has been widely used by dust cloud modelers \citep{d94, lz99, mmm02, w02, dm05}.

\begin{figure}
\begin{center}
\includegraphics[width=6.5in]{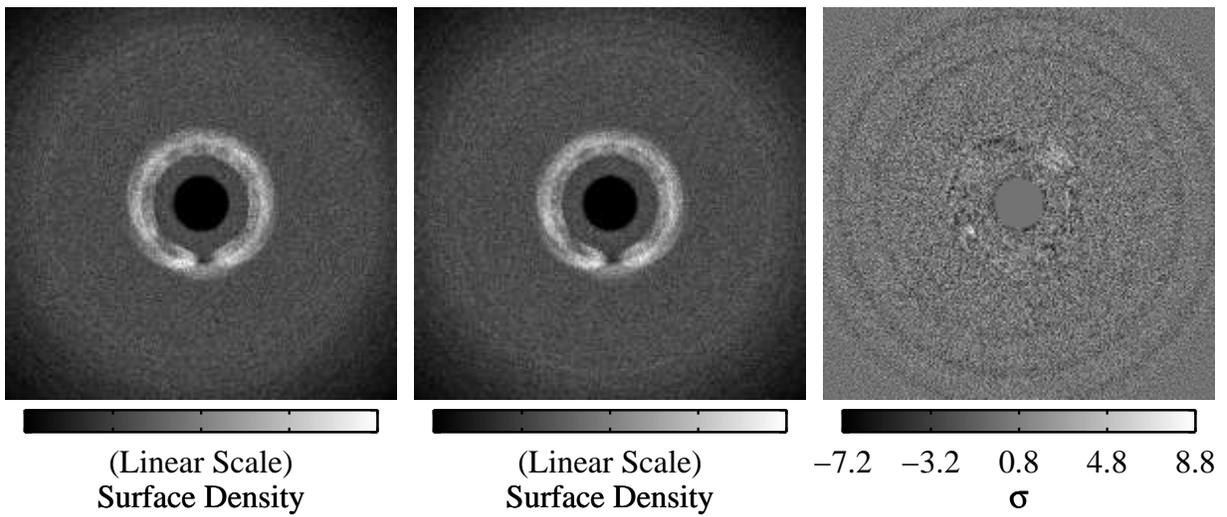}
\caption{Comparison of our hybrid symplectic integrator with a Bulirsch-Stoer integrator.  \textit{Left:} surface density histogram for 1,000 particles in the Sun-Earth system using a Bulirsch-Stoer integrator.  \textit{Middle:} surface density histrogram for the same initial conditions using our hybrid symplectic integrator.  \textit{Right:} Bulirsch-Stoer histogram minus the hybrid symplectic histogram (image is in units of $\sigma$, the $\sqrt{n}$ Poisson noise associated with the histograms).  Except in a handful of pixels, the difference is roughly consistent with Poisson noise. \label{bsvshybrid}}
\end{center}
\end{figure}

Figure \ref{bsvshybrid} shows two histograms, one for each integrator, for simulations of 1,000 particles each in the presence of the Sun/Earth system.  We used a histogram bin width of $0.0175$ AU.  For these simulations we chose $\beta=0.02$ and initially released the particles with semi-major axis, $a_{\rm dust}$, distributed uniformly between 3 and 5 AU, eccentricity, $e$, uniformly distributed from $0.0$ to $0.1$, and inclination, $i$, uniformly distributed between $0^{\circ}$ and $6^{\circ}$.  We used a symplectic time step of 0.02 years and recorded the particle locations every 250 years.

Except for a small number of pixels, the middle panel of the figure (simulation using the hybrid integrator) looks qualitatively very similar to the left-most panel (simulation using the Bulirsch-Stoer integrator).  The right panel of Figure \ref{bsvshybrid} shows the difference of these two images divided by the $\sqrt{n}$ Poisson noise expected for each pixel where $n$ is the number of particles in the pixel.  This figure demonstrates that the differences between the two models are nearly consistent with the Poisson noise of the histograms.  The two integrators resulted in histograms with minor structural differences, but the hybrid symplectic integrator runs a few times faster.

Besides pixel-to-pixel Poisson noise in the histogram, this method is also sensitive to noise in the population of MMRs.  \citetalias{mmm02} showed that the population of the dominant resonances varied by a factor of $\sim 3$ among sets of 100-particle simulations of Kuiper Belt dust interacting with Neptune.  This noise probably causes the differences between the two simulations shown in Figure \ref{bsvshybrid} beyond those attributable to pixel-to-pixel Poisson noise.  Although simulations of 100 particles may acquaint us with the generic geometry of debris disk structures, we cannot use them to predict ring contrasts; to model the contrast in a resonant cloud feature we must include enough particles to accurately populate the MMRs.

\begin{figure}
\begin{center}
\includegraphics[width=6.5in]{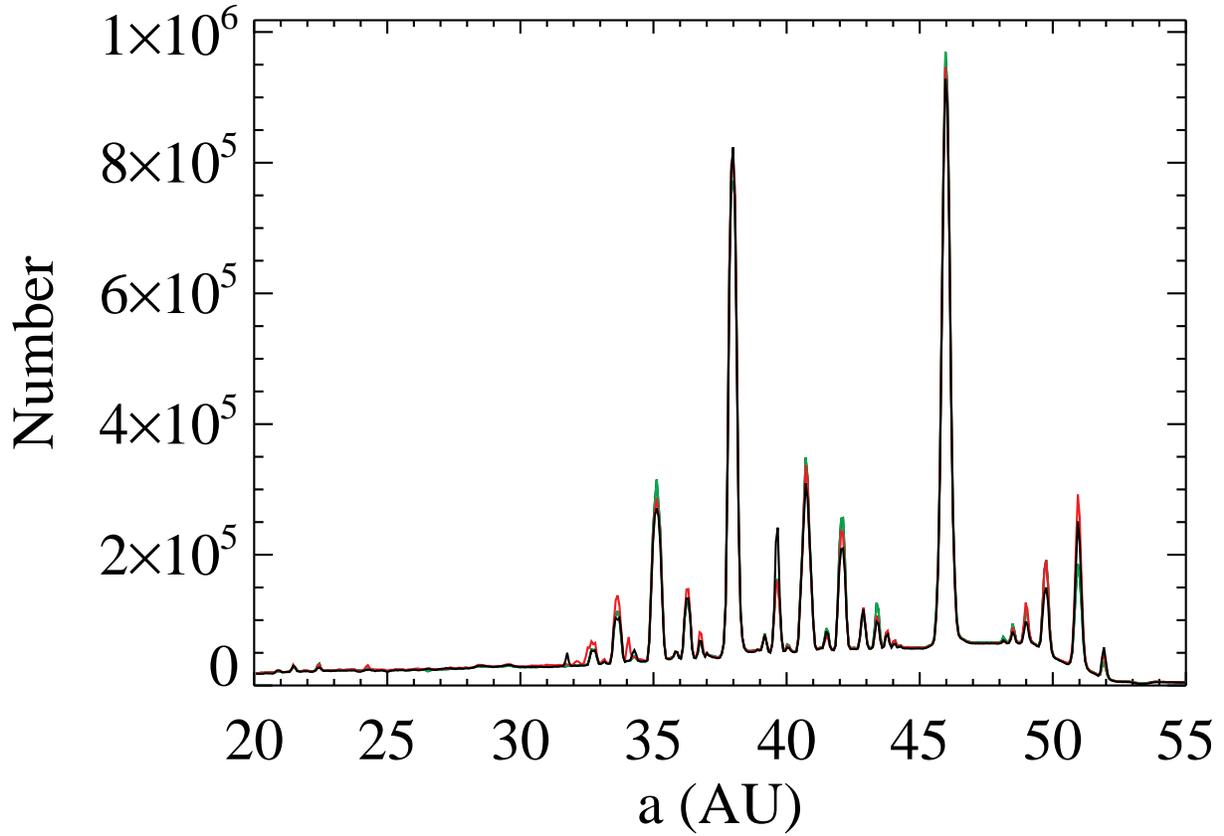}
\caption{Population of Neptune's MMRs for three independent simulations of 5,000 particles each (shown in green, red, and black).  Populations of the 2:1 and 3:2 resonances differ among the three simulations by 6.4\% and 4.3\%, respectively \citep[c.f.][Fig. 5]{mmm02}. \label{mmrs}}
\end{center}
\end{figure}

We solved this problem by using more particles.  We used the 420-processor Thunderhead cluster at NASA Goddard Space Flight Center to perform simulations of 5,000 particles each.  Figure \ref{mmrs} shows the population of MMRs for three independent 5,000-particle simulations of the Sun and four outer planets using the same initial conditions as \citetalias{mmm02}.  Simulating 5,000 particles reduced the difference between MMR populations for the three simulations to less than 7\% for the dominant 2:1 and 3:2 MMRs, allowing us to synthesize high-fidelity images and quantitatively study the resonant ring structures.

\section{Simulations \& Results}

\subsection{Cataloging Debris Disk Structure}

To explore the range of different types of structures formed by terrestrial-mass planets, we performed 120 simulations of dust interacting with single planets on circular orbits.  The simulations used 5,000 particles each and covered six values of planet mass, $M_{\rm p}$ (0.1, 0.25, 0.5, 1.0, 2.0, and 5.0 $M_{\rm \earth}$), four values of planet semi-major axis, $a_{\rm p}$ (1, 3, 6, and 10 AU), and five values of $\beta$ (0.0023, 0.0073, 0.023, 0.073, and 0.23) corresponding to spherical silicate particles ranging in radius from $\sim 1 - 120 {\rm \mu m}$.  We released the particles on orbits with semi-major axes uniformly distributed between 3.5 and 4.5 times the semi-major axis of the planet's orbit---well outside of the strongest MMRs.  We used initial eccentricities uniformly distributed between 0 and 0.2, initial inclinations uniformly distributed between 0 and $20^{\circ}$, and the longitude of the ascending node, $\Omega$, and the argument of pericenter, $\omega$, uniformly distributed between 0 and 2$\pi$.  We considered planet semi-major axes of 1 to 10 AU because typical designs for TPF can detect an exozodiacal cloud with 10 times the optical depth of the solar zodiacal cloud over roughly that range of circumstellar radii \citep{lsk06}.

We chose these initial conditions to model only dynamically-cold dust, i.e. $e_{\rm dust} \lesssim 0.2$ and $i_{\rm dust} \lesssim 20^\circ$, since this component of a dust cloud is the dominant contributor to resonant ring structure.  We neglect dynamically hot dust with the idea that it can always be added in later as a smooth background \citep{mkh04}.  The asteroid belt probably produces much of the solar system's dynamically cold dust, while comets are thought to contribute a more dynamically hot cloud component \citep[e.g.][]{ldx95, i07}.  We treat only steady-state dust clouds, assuming dust is continually replenished, and ignore transient collisional events.

\begin{figure}
\begin{center}
\includegraphics[width=6.5in]{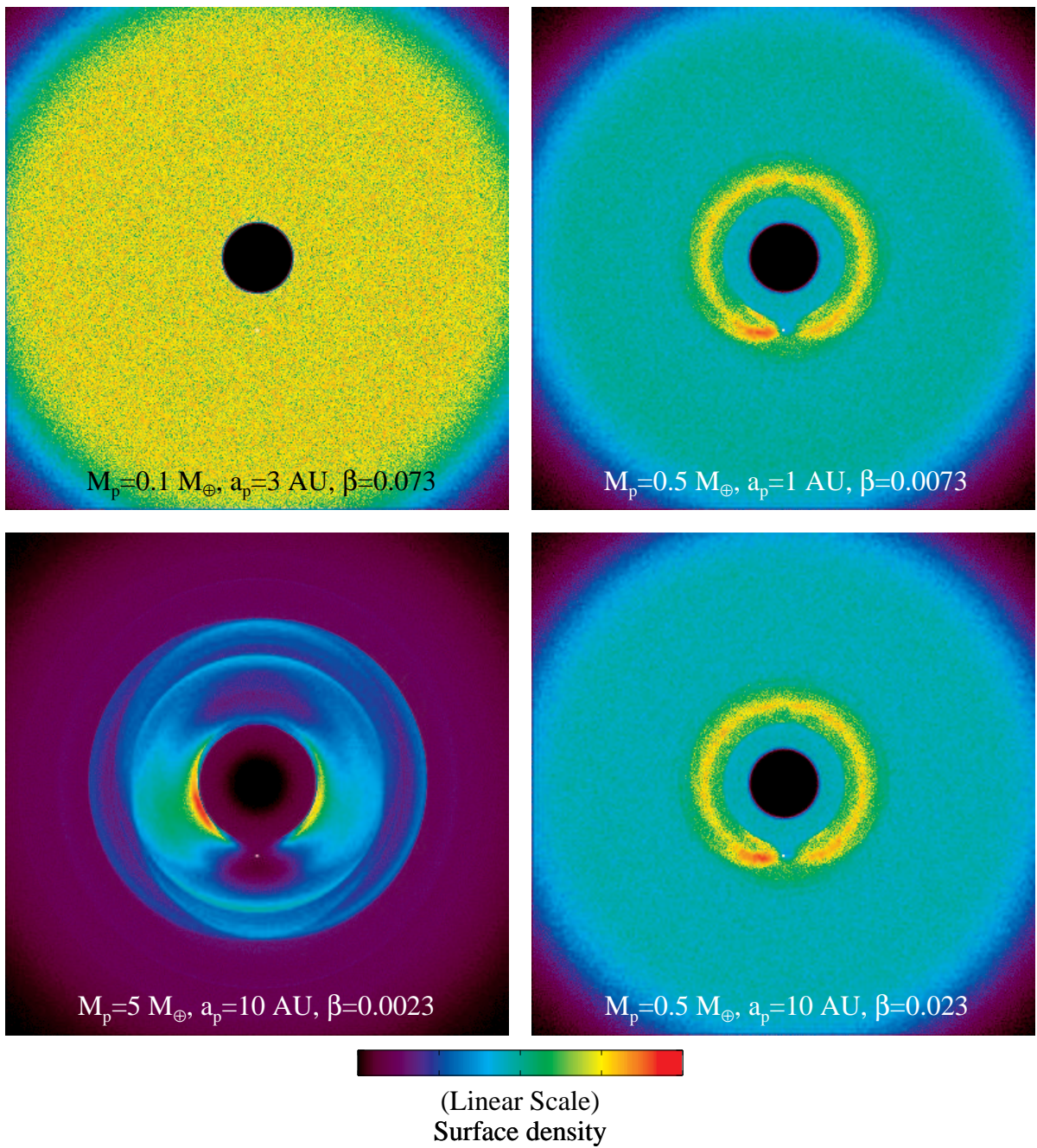}
\caption{Surface density distributions for four of the 120 simulations (scale is relative).  The star is located at the center of the image and the planet is marked with a white dot.  The planet orbits counter-clockwise in these images.  Integrations were truncated at half the planet's semi-major axis.  The simulations shown on the right have different values of $a_{\rm p}$ and $\beta$, but the same value of $\sqrt{a_{\rm p}}/\beta$.  Their surface density distributions are nearly identical; their difference is consistent with Poisson noise (see Section \ref{ringcontrast}).\label{densities}}
\end{center}
\end{figure}

Figure \ref{densities} shows some examples of the histograms from our simulations, which reveal a wide range of trapping behavior.  Some histograms show no azimuthal or radial structure, while others show high contrast rings.  All of the patterns are Type I structures as identified by \citet{kh03}.

Several general trends emerged.  The ring contrast increased with increasing planet mass.  Reducing $\beta$ also enhanced trapping, as did increasing the planet's semi-major axis.

These last two trends can be explained by comparing the libration time of a given MMR to the PR time.  The PR time scales as $a_{\rm dust}^2 / \beta$, while the libration time for a given resonance scales as $a_{\rm dust}^{3/2}$, where $a_{\rm dust}$ is the semi-major axis of a dust grain's orbit.  The ratio of these quantities yields $\sqrt{a_{\rm dust}} / \beta$, a parameter that measures the degree to which resonant trapping is adiabatic \citep[e.g.][]{h82}; the trapping becomes more adiabatic and more efficient at greater distances from the star, and for larger particles.  We discuss this phenomenon further in Section \ref{adiabaticity} below.

\begin{figure}
\begin{center}
\includegraphics[width=6.5in]{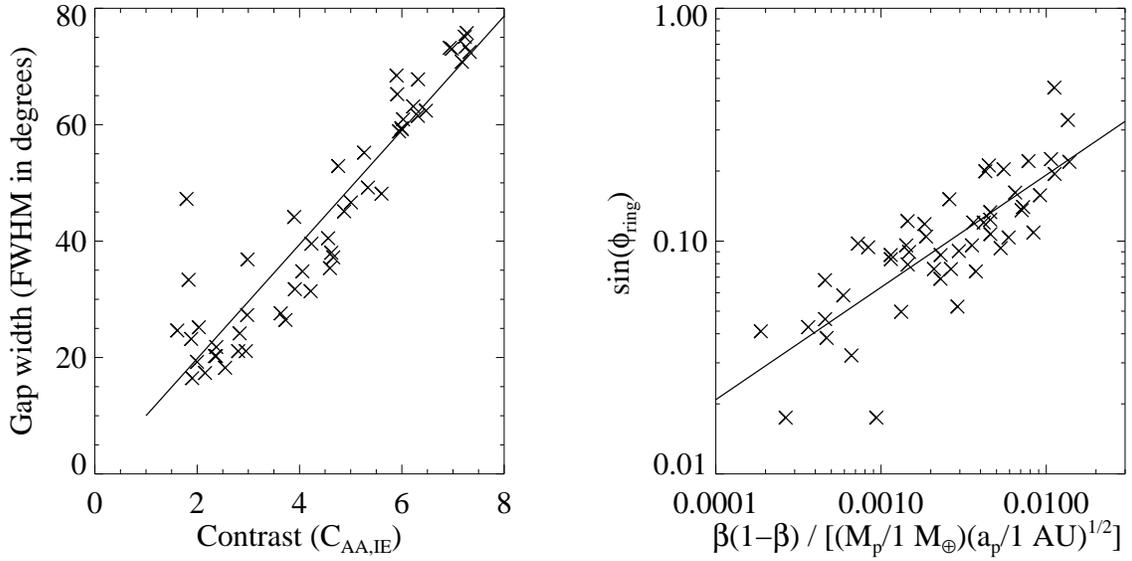}
\caption{\textit{Left:} the angular size of the ``gap" in the ring structure around the location of the planet in our simulations versus the contrast of the ring structure.  \textit{Right:} the sine of the prograde shift plotted against the function $\beta(1-\beta)/(M_{\rm p} a_{\rm p}^{1/2})$.  We removed all data with $C_{\rm AA,IE} < 1.6$ from these plots.  Solid lines show linear fits to the data. \label{gapfigure}}
\end{center}
\end{figure}

In addition to following these trends, all of the simulated ring structures, like those shown in Figure \ref{densities}, share some salient features:
\begin{enumerate}
	\item For cases in which even a modest amount of trapping occurs (azimuthally averaged contrasts $\gappeq 1.3:1$), the ring structures exhibit a sharp inner edge at $\approx 0.83 a_{\rm p}$.  This feature probably appears because the eccentricities of particles trapped in exterior MMRs are typically pumped up to a limiting value before a close encounter with the planet ejects them from resonance.  For a particular MMR, all particles, regardless of $\beta$, tend to approach the same limiting eccentricity and accordingly, a similar pericenter distance \citep{bfm94}.  The limiting eccentricities are such that the limiting pericenter distances are nearly equal for the dominant resonances (e.g. the 2:1 and 3:2 resonances have limiting pericenter distances of $0.823a_{\rm p}$ and $0.827a_{\rm p}$, respectively), creating the ring structure's sharp inner edge.
	\item A gap in the ring structure, a local minimum in the surface density, appears around the planet.  If we define gap width as the FWHM of the minimum in the azimuthal surface density profile at $r=a_{\rm p}$, we find that the gap width is linearly proportional to the contrast of the ring, as shown in the left panel of Figure \ref{gapfigure}.  A linear fit to the data shown in this figure gives $w_{\rm gap} \approx 10^\circ \times C_{\rm AA,IE}$ for $C_{\rm AA,IE} > 1.6$, where $w_{\rm gap}$ is the gap width in degrees and $C_{\rm AA,IE}$ is the azimuthally averaged inner-edge contrast (see Section \ref{ringcontrast}).
	\item The rings show a leading-trailing asymmetry.  The trailing side of the ring structure is noticeably denser than the leading side, and the structure is rotationally shifted in the prograde direction causing the trailing side to be closer to the planet \citep{d94}.  To examine the leading-trailing asymmetry caused by a prograde shift of the ring structure, we measured the azimuthal offset of the center of the gap described above from the planet.  The right panel of Figure \ref{gapfigure} shows these measured prograde shifts of our simulations.  \citet{kh03} showed for a particular first order exterior MMR,
\begin{equation}
	\label{progradeshift}
	\sin{\phi_0} \propto \frac{\beta\left(1-\beta\right)}{ M_{\rm p} \sqrt{a_{\rm p}}},
\end{equation}
where $\phi_0$ is the prograde shift of the pericenter.  Therefore, we plotted the sine of each measured prograde shift against $\beta(1-\beta)/(M_{\rm p}\sqrt{a_{\rm p}})$ in the right panel of Figure \ref{gapfigure}.  Our data reveal the approximate proportionality
\begin{equation}
	\label{progradeshift2}
	\sin{\phi_{\rm ring}} \propto \left[\frac{\beta\left(1-\beta\right)}{ M_{\rm p} \sqrt{a_{\rm p}}}\right]^{0.5},
\end{equation}
where $\phi_{\rm ring}$ is the measured prograde shift of the ring structure.  While the relationship in Equation \ref{progradeshift} holds for a single MMR, it does not strictly apply to a given ring structure which consists of several well-populated MMRs.  The relative populations of these MMRs are also functions of $M_{\rm p}$, $a_{\rm p}$, and $\beta$.  However, this situation seems to preserve a power-law relationship between $\sin{\phi_{\rm ring}}$ and $\beta\left(1-\beta\right) / \left(M_{\rm p} \sqrt{a_{\rm p}}\right)$, as shown in Equation \ref{progradeshift2}.

	\item The radial width of the ring increases with the contrast of the ring, ranging from a few percent of $a_{\rm p}$ to $\sim 1.6a_{\rm p}$ in the highest contrast case.  As the trapping probabilities of all the MMRs increase, MMRs farther from the planet's orbit become populated.  For this reason, the outer-edge of the ring structure differs significantly among simulations.  The outer-edge can be quite blurry or very well-defined, making the radial width of a ring structure difficult to quantify.

\end{enumerate}

Our catalog of debris disk structures induced by terrestrial-mass planets is publicly available online at http://asd.gsfc.nasa.gov/Christopher.Stark/catalog.php.  This online catalog also contains images synthesized from the density distributions in scattered light and 10 $\mu$m thermal emission assuming blackbody grains.  Future studies of resonant ring structures with TPF or other experiments can use our catalog to interpret dust cloud patterns in terms of planet and dust parameters, assuming the observed image is dominated by a single grain size.  We envision the following process, inspired by recent papers on disks observed with the Hubble Space Telescope \citep[e.g.][]{c03, kgc05}:

\begin{enumerate}
	\item Deproject the image to remove inclination effects.
	\item Remove any smooth backgrounds by a power law fit.
	\item Estimate the dominant grain size in the resonant ring using infrared photometry or other methods.
	\item Compare the image of the disk to the online catalog to constrain the planet's mass and location.
\end{enumerate}

\subsection{Ring Contrast \label{ringcontrast}}

We considered three different metrics for describing the ring contrast in our simulations:

$C_{\rm Max}$: The surface density of the ring at its densest point divided by the surface density of the background cloud

$C_{\rm AA,Max}$: The maximum value of the azimuthally averaged surface density divided by the surface density of the background cloud

$C_{\rm AA,IE}$: The azimuthally averaged surface density at the inner edge of the ring divided by the surface density of the background cloud

We calculated the above contrast metrics for all 120 simulations.  We measured the surface density of the background cloud at a circumstellar distance $r \approx 0.8 a_{\rm p}$.  The surface density of the background cloud was nearly constant inside and outside of the ring, but did exhibit a small local minimum near $r \approx 0.8 a_{\rm p}$ in a few cases.

To calculate $C_{\rm Max}$, we must search for the densest pixel, which introduces a bias toward pixels that exhibit an extreme amount of Poisson noise.  To reduce this noise we averaged the surface density over nine pixels centered on the densest point.  Using $C_{\rm AA,Max}$ or $C_{\rm AA,IE}$, on the other hand, automatically averages over the effects of Poisson noise in our simulations.

\begin{figure}
\begin{center}
\includegraphics[width=6.5in]{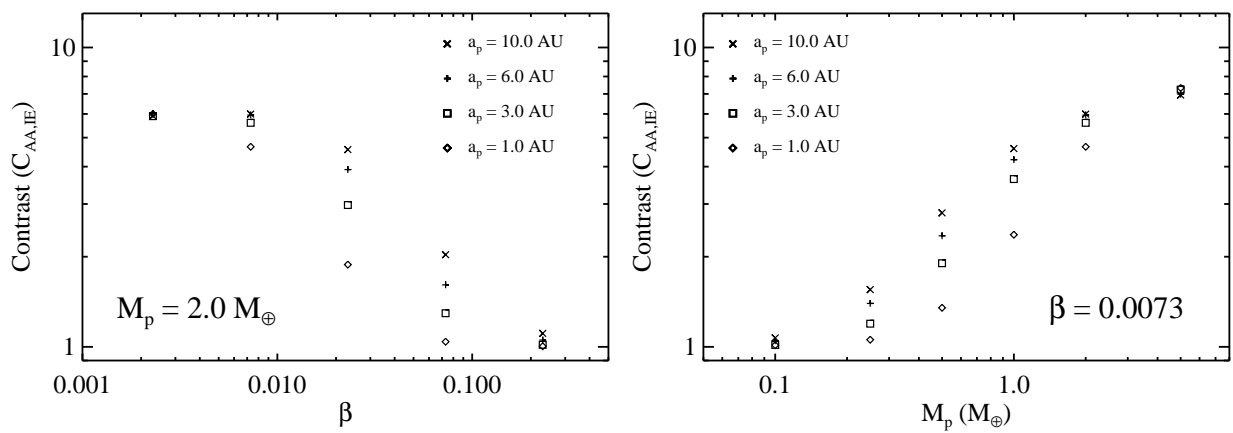}
\caption{The azimuthally-averaged contrast measured at the inner edge of the ring structure (see Section \ref{ringcontrast} for definition of contrast) as a function of $\beta$ (left panel) and planet mass (right panel).  Both figures show a transition from a no-trapping regime to a saturation regime where contrast is independent of semi-major axis. \label{contrastvsparams}}
\end{center}
\end{figure}

Figure \ref{contrastvsparams} shows two examples of how the contrast, $C_{\rm AA,IE}$, depends on planet mass and $\beta$.  Both plots show a similar behavior with three distinct regions: a no-trapping regime (contrast $\sim$1), a transitional regime, and a saturation regime (maximum contrast).  The saturation regime is of particular significance.  Our results suggest that within the range of parameters investigated, for a given value of $\beta$, all contrasts converge to the same value for large planet masses independent of planet semi-major axis, i.e. the contrast becomes ``saturated" and increasing the planet's semi-major axis has little effect on the contrast.  The right panel in Figure \ref{contrastvsparams} illustrates this behavior; all four contrast curves, each of which corresponds to a different planet semi-major axis, approach the same value of $\sim 7$ near $M_{\rm p} = 5\; M_{\earth}$. Similarly, for a given planet mass, contrasts converge to the same value for small $\beta$ independent of $a_{\rm p}$, as shown in the left panel in Figure \ref{contrastvsparams}.  The morphology of the structure can vary, but the contrast of the ring structure is roughly constant in these saturation regimes.

\subsection{Adiabaticity\label{adiabaticity}}

As we mentioned above, dividing the PR time by the libration time of a given MMR yields a parameter, $\sqrt{a_{\rm p}}/\beta$, that indicates the degree to which the resonant trapping is adiabatic.  We plot the contrast in our simulations as a function of this parameter in Figure \ref{contrasttrends}.  This figure demonstrates that for $M_{\rm p} \lesssim 5\; M_{\earth}$ and for a given distribution of parent body orbital elements, the ring contrast is a function of only two parameters: planet mass and $\sqrt{a_{\rm p}}/\beta$.

The morphology of the resonant rings is also, to good approximation, a function of only the planet mass and $\sqrt{a_{\rm p}}/\beta$.  The models shown in the two right panels of Figure \ref{densities} illustrate this phenomenon; for both models $\sqrt{a_{\rm p}}/\beta \approx 137\; {\rm AU}^{1/2}$.  These two models have the same morphology to a level consistent with pixel-to-pixel Poisson noise.  Note that for small $\beta$ in Equation \ref{progradeshift2}, the prograde shift is approximately a function of $\left(\sqrt{a_{\rm p}}/\beta\right)^{-1}$ for a given planet mass.

For large values of $\beta$ and $M_{\rm p}$, the morphology and contrast of the ring structures are not simple functions of $\sqrt{a_{\rm p}}/\beta$.  Simulations with large values of $\beta$, but equal values of $\sqrt{a_{\rm p}}/\beta$ (e.g. $\sqrt{1\; {\rm AU}}/0.073$ and $\sqrt{10\; {\rm AU}}/0.23$) show morphological differences, including differences in prograde shift.  Our simulations with $M_{\rm p} = 5\; M_{\earth}$ also show contrast differences among rings with equal values of $\sqrt{a_{\rm p}}/\beta$.

\citet{w03} investigated resonant trapping in MMRs for a system of planetesimals exterior to an outward migrating planet on a circular orbit.  \citet{w03} plotted the trapping probability for a single MMR in his model as a function of migration rate and planet mass and found it could be well approximated by a function of the form $P = [1+(\dot{a}_{\rm p}/p_1)^{p_2}]^{-1}$, where $\dot{a}_{\rm p}$ is the migration rate and the parameters $p_1$ and $p_2$ are power laws in planet mass.  Our trapping scenario assumes dust migrating inward toward the planet, but the concept is similar.  Since contrast is closely related to trapping probability, we decided to fit the data shown in Figure \ref{contrasttrends} with a function of the form
\begin{equation}
 	C = 1 + p_1 \left(1 + \left( \frac{p_2}{\sqrt{a_{\rm p}} / \beta} \right)^{p_3} \right)^{-1},
	 \label{contrastequation}
\end{equation}
inspired by \citet{w03}.  Each of the three parameters, $p_i$, is a power law in planet mass of the form $p_i = p_{i,1}M_{\rm p}^{p_{i,2}}$.  We fit all 120 contrast measurements with this six-parameter function for each of the three contrast metrics.  The best fits, two of which are shown in Figure \ref{contrasttrends}, are:

$C_{\rm AA,IE}$: $p_1 \approx 4.38\left(M_{\rm p}/M_{\earth}\right)^{0.19}$, $p_2 \approx 207\left(M_{\rm p}/M_{\earth}\right)^{-1.17}$, $p_3 \approx 2.05\left(M_{\rm p}/M_{\earth}\right)^{0.11}$

$C_{\rm AA,Max}$: $p_1 \approx 4.54\left(M_{\rm p}/M_{\earth}\right)^{0.17}$, $p_2 \approx 205\left(M_{\rm p}/M_{\earth}\right)^{-1.17}$, $p_3 \approx 1.63\left(M_{\rm p}/M_{\earth}\right)^{0.19}$

$C_{\rm Max}$: $p_1 \approx 6.23\left(M_{\rm p}/M_{\earth}\right)^{0.27}$, $p_2 \approx 164\left(M_{\rm p}/M_{\earth}\right)^{-1.09}$, $p_3 \approx 1.72\left(M_{\rm p}/M_{\earth}\right)^{0.05}$

Equation \ref{contrastequation}, combined with the above values, summarizes our results for all combinations of planet mass, planet semi-major axis and $\beta$ we simulated.  Figure \ref{contrasttrends} shows the inner-edge contrast, $C_{\rm AA,IE}$, deviates significantly from the fits for large $M_{\rm p}$ and large $\sqrt{a_{\rm p}}/\beta$.  The increased trapping efficiency for MMRs with these massive planets likely enhances the population of MMRs farther from the planet's orbit and depletes the inner MMRs that cause the sharp inner edge.

\section{Multi-Particle-Size Models \label{dohnanyisection}}

\subsection{Composite Simulations \label{compositesimulations}}

We used our 120 simulations to produce 20 multiple-particle-size dust cloud models by forming weighted sums of the histograms assuming a Dohnanyi distribution of particle sizes \citep{d69}.  Each of these composite models effectively utilizes 25,000 particles.  Exactly how we apply the ideas in \citet{d69} has profound effects on our composite models, so we present here two different kinds of models.

First, we assembled a composite model in which the particles are initially released from their parent bodies according to a crushing law, and do not undergo any further collisional processing as they spiral inward.  This scenario models a sparse disk with a belt of dust-producing material, like our own zodiacal cloud.  The crushing law for asteroid material at micron sizes is unknown, so we choose the crushing law used by \citet{d69}:
\begin{equation}
\label{crushinglaw}
\frac{dN}{ds} \propto s^{-\alpha},
\end{equation}
where $dN$ is the number of particles with radius $s$ in a bin of width $ds$, and $\alpha = 3.4$.

We calculate the optical depth, $\tau$, for our composite models from $\tau = \sum_i w_i A_i \sigma_i$, where $w_i$, $A_i$, and $\sigma_i$ are the weighting factor, particle cross-section, and surface number density of the $i^{\rm th}$ single-particle simulation, respectively.  We assume that the cross-section of each particle is $A_i \propto \beta^{-2}$.  The crushing law in Equation \ref{crushinglaw} implies a weighting factor for the $i^{\rm th}$ histogram of $w_i = \beta_i^{\alpha-2} \Delta\beta_i$, where $\Delta\beta_i$ is the width of the $\beta_i$ bin.  For a constant logarithmic spacing in $\beta$, like the spacing we used in our simulations, and the Dohnanyi crushing law, $w_i = \beta_i^{2.4}$.

Larger particles have longer PR times, so in the absence of collisional processing, their density is enhanced by a factor of $\beta^{-1}$ under our assumption of a steady-state cloud model.  One might expect that this effect must be included in the weighting factor.  However, our simulations include this effect automatically as long as we keep the frequency with which particle locations are recorded constant among all of our simulations.  We did, in fact, vary the recording frequency with the PR time, but we corrected for the differences in recording frequency before summing the histograms.

\begin{figure}
\begin{center}
\includegraphics[width=6.5in]{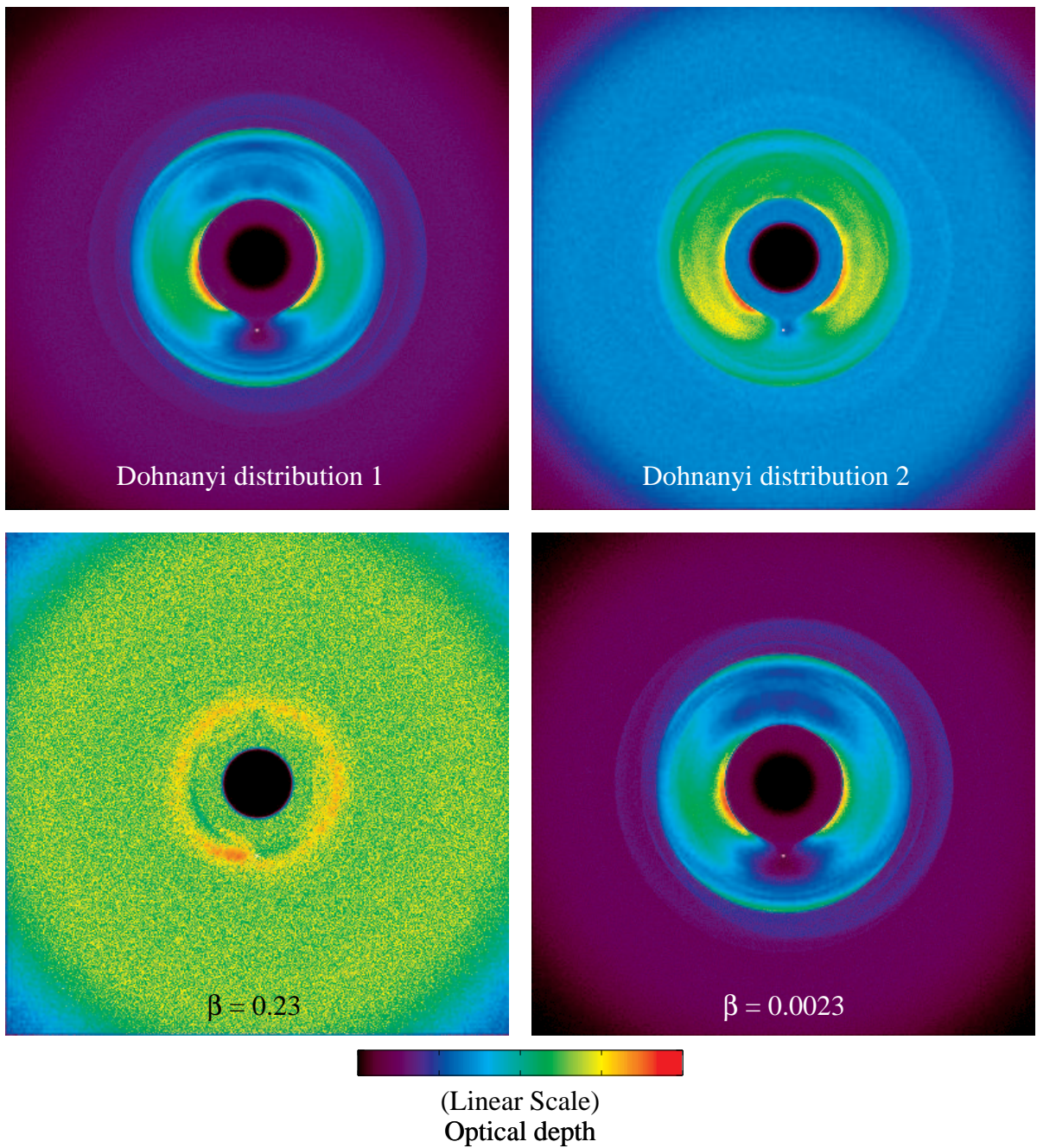}
\caption{Comparison of the optical depths for a composite cloud formed by two different methods for $M_{\rm p} = 2.0\; M_{\earth}$ and $a_{\rm p} = 6.0$ AU. The planet, marked with a white dot, orbits counter-clockwise in these images.  \textit{Top-left:} A composite collisionless cloud where the particles are released with a size distribution equal to the crushing law used by \citet{d69}.  \textit{Top-right:} The same composite cloud, but formed by forcing the surface density outside of the ring structure to obey a Dohnanyi distribution.  \textit{Bottom-left:} The optical depth of the smallest particles included in the composite clouds.  \textit{Bottom-right:} The optical depth of the largest particles included in the composite clouds.  The largest particles dominate the optical depth in a cloud of particles released with a Dohnanyi crushing law. \label{dohnanyicomparison}}
\end{center}
\end{figure}

Figure \ref{dohnanyicomparison} shows the optical depth of one of our 20 composite models ($M_{\rm p} = 2.0\; M_{\earth}$, $a_{\rm p} = 6.0$ AU), together with the optical depths of single-particle-size models using only the smallest and largest particle sizes included in the composite model.  Although the crushing law used by \citet{d69} favors smaller particles by number, even more than some empirical crushing laws \citep{d07}, the optical depth in the composite models is dominated by the largest particles.  This situation occurs because the larger particles are both longer lived ($t_{\rm PR} \propto\beta^{-1}$) and more likely to be trapped in MMRs.  Hence, the upper left panel of the figure closely resembles the lower right panel.

Next, for the purpose of illustration, we ignored the initial size distribution of dust particles and forced the disk to obey a size distribution of 
\begin{equation}
\frac{dN}{ds} \propto s^{-3.5}
\end{equation}
at a radius of $\sim 3 a_{\rm p}$ from the star.  This scenario probably doesn't have a physical interpretation, but it illustrates an interesting phenomenon: how resonant trapping tends to sort particles by size.  We enforce the size distribution at one location within the disk, but the size distribution will not follow a Dohnanyi distribution elsewhere in the disk.

The top right panel in Figure \ref{dohnanyicomparison} shows the optical depth of an example of this kind of composite cloud, normalized to a Dohnanyi distribution at $\sim 18$ AU.  Models constructed in this fashion are more greatly affected by the smallest grains.  Hence, the top right panel does not greatly resemble the lower right panel in Figure \ref{dohnanyicomparison}.

\subsection{Semi-Analytic Treatment}

We can further develop these ideas with a simple semi-analytic treatment.  For a given planet mass and semi-major axis, the contrast function (see Equation \ref{contrastequation}) becomes $C(s)$, where $s$ is the particle size.  We approximate the contrast function in Equation \ref{contrastequation} with the piecewise function
\begin{equation}
	\label{piecewisefunction}
	C(s) = \left\{ \begin{array}{ll}
	1 & \mbox{for $s < s_1$} \\
	\left(\frac{s}{s_1}\right)^m & \mbox{for $s_1 < s < s_2$} \\
	C_{\rm large} & \mbox{for $s > s_2$},
	\end{array} \right.
\end{equation}
where $C_{\rm large} = 1 + p_1$ is the contrast for the largest particles, $m$ is the logarithmic slope of the contrast in the transition regime, and $s_1$ and $s_2$ are the particle sizes that mark the beginning and end of the transition regime, respectively.  We fit our contrast data with this piecewise function and obtained the following power law estimates assuming silicate grains ($\rho \sim 2\; {\rm g\; cm^{-3}}$):
\begin{equation}
	\label{clargeequation}
	C_{\rm large; AA, IE} \approx 1 + 4.38 \left(\frac{M_{\rm p}}{M_\earth}\right)^{0.19}
\end{equation}
\begin{equation}
	m \approx 0.6 \left(\frac{M_{\rm p}}{M_\earth}\right)^{0.18}
\end{equation}
\begin{equation}
	\label{s1equation}
	\left(\frac{s_1}{1\; \mu{\rm m}}\right) \approx 10 \left(\frac{M_{\rm p}}{M_\earth}\right)^{-1.12} \left(\frac{a_{\rm p}}{1\; {\rm AU}}\right)^{-0.5}
\end{equation}
\begin{equation}
	\label{s2equation}
	\left(\frac{s_2}{1\; \mu{\rm m}}\right) \approx 150 \left(\frac{M_{\rm p}}{M_\earth}\right)^{-1.35} \left(\frac{a_{\rm p}}{1\; {\rm AU}}\right)^{-0.5}
\end{equation}

In the same manner as Section \ref{ringcontrast}, we defined the contrast of any ring structure as the surface density within the ring, $\sigma_{\rm ring}$, divided by the background surface density, $\sigma_{\rm BG}$.  The contrast in optical depth of a cloud containing several components of various sized particles, labeled with the index $i$, is
\begin{equation}
\langle C_\tau \rangle = \frac{\displaystyle\sum_i C_i \sigma_{{\rm BG,}i} A_i}{\displaystyle\sum_i \sigma_{{\rm BG,}i} A_i},
\end{equation}
For a collisionless cloud with a continuous distribution of grain sizes, the contrast in optical depth of the composite cloud is given by
\begin{equation}
	\label{ctauintegral}
	\langle C_{\tau} \rangle = \frac{\displaystyle\int_{s_{\rm min}}^{s_{\rm max}} s^{3-\alpha} C(s) \,ds}{\displaystyle\int_{s_{\rm min}}^{s_{\rm max}}s^{3-\alpha}\,ds},
\end{equation}
where we have explicitly included the particle cross section ($A(s) \propto s^2$) and background surface density ($\sigma_{\rm BG}(s) \propto s^{1-\alpha}$).  For a collisionless cloud, the background surface density is enhanced by a factor of $s$ due to the PR time scaling as $s$ (see Section \ref{compositesimulations}), and a factor of $s^{-\alpha}$, which describes the assumed crushing law.  

Using Equations \ref{piecewisefunction}, we can now integrate Equation \ref{ctauintegral} directly.  Assuming $s_{\rm min}<s_1<s_2<s_{\rm max}$ and $(s_{\rm min} / s_{\rm max})^{|4-\alpha|} \ll 1$ when $\alpha \ne 4$, we find
\begin{equation}
	\label{analyticcontrast}
	\langle C_{\tau} \rangle = \left\{ \begin{array}{ll}
		C_{\rm large} - \left(\frac{s_2}{s_{\rm max}}\right)^{4-\alpha} \left[ C_{\rm large} - \frac{4-\alpha}{4-\alpha + m} \left( \frac{s_2}{s_1} \right)^m \right] +  \frac{m}{4-\alpha+m}\left( \frac{s_1}{s_{\rm max}} \right)^{4-\alpha} & \mbox{for $\alpha < 4$} \\
		\left(\ln \left(\frac{s_1}{s_{\rm min}}\right) + C_{\rm large}\ln \left(\frac{s_{\rm max}}{s_2}\right) + m^{-1} \left[\left(\frac{s_2}{s_1}\right)^m - 1\right]\right)/\ln \left(\frac{s_{\rm max}}{s_{\rm min}}\right) & \mbox{for $\alpha = 4$} \\
		1 + \left(\frac{s_2}{s_{\rm min}}\right)^{4-\alpha} \left[ C_{\rm large} - \frac{4-\alpha}{4-\alpha+m}\left(\frac{s_2}{s_1}\right)^m\right] - \frac{m}{4-\alpha+m}\left(\frac{s_1}{s_{\rm min}}\right)^{4-\alpha} & \mbox{for $\alpha > 4$, $\alpha \neq 4+m$} \\
		1 + C_{\rm large}\left(\frac{s_2}{s_{\rm min}}\right)^{4-\alpha} - \left(\frac{s_1}{s_{\rm min}}\right)^{4-\alpha}\left(1+\ln{\left(\frac{s_2}{s_1}\right)}\right) & \mbox{for $\alpha > 4$, $\alpha = 4+m$}.
		\end{array} \right.
\end{equation}
For cases in which the maximum particle size in a disk is less than $s_2$ (see Equation \ref{s2equation}), simply replace all instances of $s_2$ in Equations \ref{analyticcontrast} with $s_{\rm max}$.  Similarly, for cases in which the minimum particle size in a disk is greater than $s_1$ (see Equation \ref{s1equation}), replace all instances of $s_1$ with $s_{\rm min}$.

Equations \ref{analyticcontrast}, together with Equations \ref{clargeequation}--\ref{s2equation}, give analytic expressions for optical depth contrast in terms of $M_{\rm p}$, $a_{\rm p}$, $s_{\rm min}$, and $s_{\rm max}$.  Although Equations \ref{analyticcontrast} address all possible scenarios, the most plausible scenarios have crushing laws with $\alpha < 4$ \citep[e.g.][]{d07}.  With this assumption, Equations \ref{analyticcontrast} combined with Equations \ref{clargeequation}--\ref{s2equation} gives
\begin{equation}
	\label{analyticcontrast2}
	\langle C_{\tau{\rm ;AA, IE}} \rangle \approx \left\{ \begin{array}{ll}
		1 + 4.4M_{\rm p}'^{0.19} + 
			s_{\rm max}'^{\alpha-4} X \left[\left( 10M_{\rm p}'^{-1.12}a_{\rm p}'^{-0.5}\right)^{4-\alpha} \right. & \\
			\indent \left. - \left( 1 + 4.4M_{\rm p}'^{0.19}\right) \left( 150 M_{\rm p}'^{-1.35}a_{\rm p}^{-0.5}\right)^{4-\alpha} \right] & \mbox{for $s_{\rm max} > s_2$} \\

		X s_{\rm max}'^{\alpha-4} \left(  10 M_{\rm p}'^{-1.12} a_{\rm p}'^{-0.5}\right)^{4-\alpha} & \\
		\indent + \left(1-X\right) s_{\rm max}'^{0.6M_{\rm p}'^{0.18}}\left( 10 M_{\rm p}'^{-1.12} a_{\rm p}'^{-0.5}\right)^{-0.6M_{\rm p}^{0.18}}, & \mbox{for $s_{\rm max} < s_2$}
	\end{array} \right.
\end{equation}
where $M_{\rm p}' = \left(\frac{M_{\rm p}}{M_\earth}\right), a_{\rm p}' = \left(\frac{a_{\rm p}}{1\; {\rm AU}}\right), s_{\rm max}' = \left(\frac{s_{\rm max}}{1\; \mu{\rm m}}\right)$, and $X = \frac{0.6 M_{\rm p}'^{0.18}}{4-\alpha+0.6 M_{\rm p}'^{0.18}}$.

If, as in our first composite model in Section \ref{compositesimulations}, we assume that the particles are released from their parent bodies in accordance with the \citet{d69} crushing law and then spiral inward without colliding, $\alpha = 3.4$.  In this case, Equations \ref{analyticcontrast} give a contrast in optical depth of $\langle C_{\tau} \rangle \approx  C_{\rm large}$ in the limit $s_{\rm max} \gg s_2$.  This result confirms our numerical results for our first composite model, shown in the upper left panel of Figure \ref{dohnanyicomparison}; the contrast in optical depth is dominated by the large particles.

For our second composite model, we forced the background density to obey a Dohnanyi distribution at $\sim 3 a_{\rm p}$, i.e. $\sigma_{\rm BG}(s) \propto s^{-3.5}$, so that $\alpha = 4.5$.  This technique essentially removes the factor of $s$ in the background surface density that results from the PR time scaling as $s$.  For the composite cloud shown in Figure \ref{dohnanyicomparison}, $M_{\rm p} = 2.0\; M_{\earth}$, and $a_{\rm p} = 6.0$ AU, for which $m \approx 0.68$, $s_1 \approx 1.9\; \mu{\rm m}$, $s_2 \approx 24\; \mu{\rm m}$, and $C_{\rm large,AA,IE} \approx 6$.  We let each simulated particle size represent a range of particle sizes using the midpoint method, which gives $s_{\rm min} \approx 0.7\; \mu{\rm m}$.  With these values, Equations \ref{analyticcontrast} give a contrast in optical depth of $\langle C_{\tau {\rm,AA,IE}} \rangle \approx 2.4$, in agreement with the measured contrast in the top right panel of Figure \ref{dohnanyicomparison}.

\begin{figure}
\begin{center}
\includegraphics[width=6.5in]{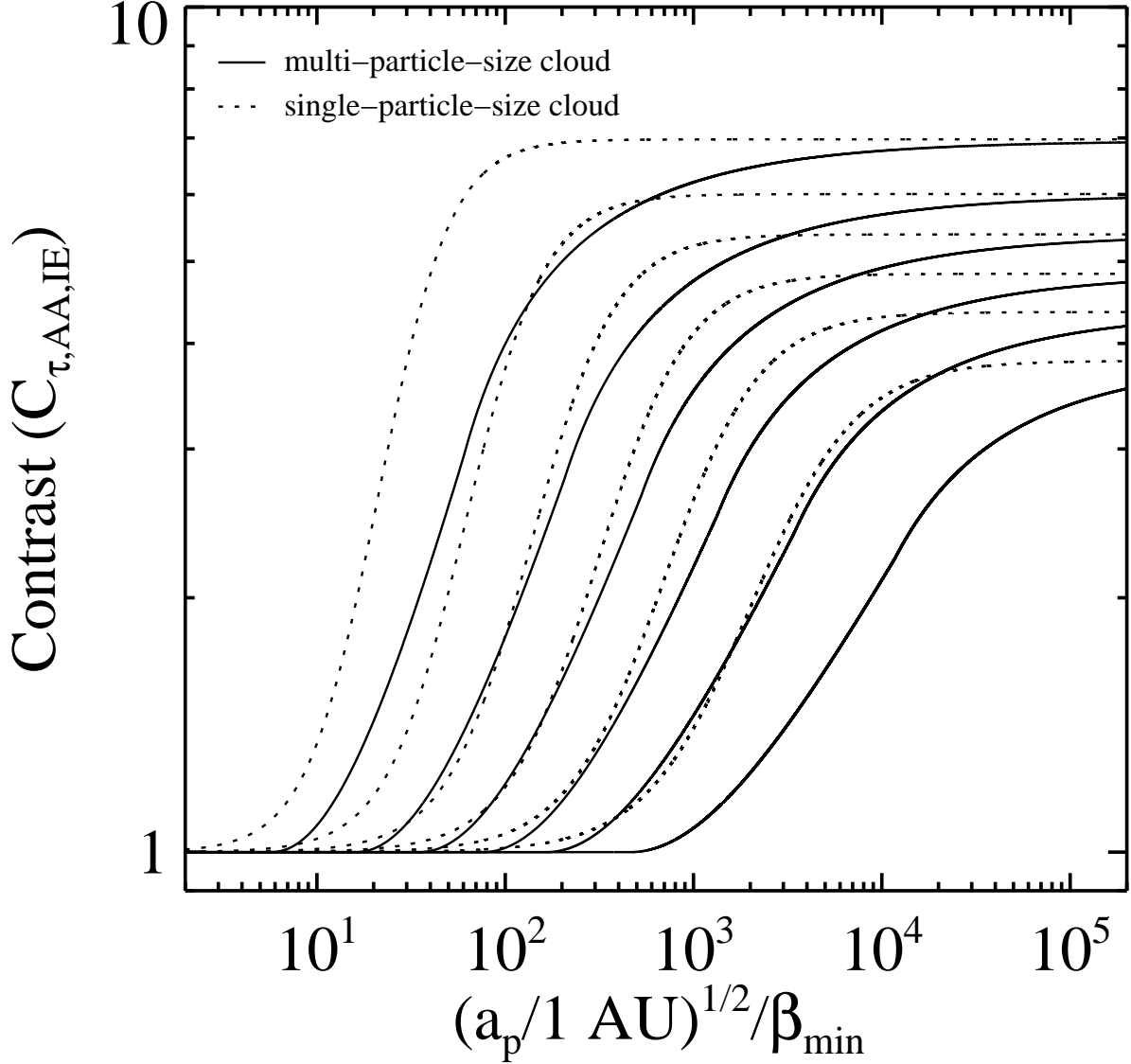}
\caption{Contrast in optical depth for multi-particle-size clouds (solid lines) compared to single-particle-size clouds (dashed lines) assuming a \citet{d69} crushing law ($\alpha = 3.4$; see Equations \ref{finalcontrastequation}).  From top to bottom, the six solid lines and six dashed lines correspond to six values of planet mass: 5, 2, 1, 0.5, 0.25, and 0.1 $M_\earth$.  The contributions of the small grains reduce the contrasts of the multi-particle-size clouds compared to single-particle-size clouds with the same minimum value of $\beta$. \label{multivssingle}}
\end{center}
\end{figure}

Figure \ref{multivssingle} illustrates in general how a distribution of particle sizes affects the contrast of a ring structure.  This figure compares the contrast of a collisionless multi-particle-size cloud (Equations \ref{analyticcontrast}) to that of a single-particle-size cloud as a function of $\sqrt{a_{\rm p}}/\beta_{\rm min}$ assuming a \citet{d69} crushing law.  Both kinds of clouds have the same contrast in the adiabatic limit (large $\sqrt{a_{\rm p}}/\beta_{\rm min}$), but the contribution of the smaller grains reduces the contrast elsewhere, effectively broadening the transition between the no-trapping regime and saturation regime.  Crushing laws with $\alpha < 3.4$ result in contrast curves that more closely resemble the single-particle-size contrast curves shown in Figure \ref{multivssingle}.

In a real zodiacal cloud, collisions affect the distribution of grains, even far from the source of the grains.  Our composite dust cloud models do not include collisions and become unreliable for particles with collisional times less than their PR times.  Our composite models also lack the structural results of collisional effects, such as the loss of particles as a function of circumstellar distance \citep{w05} and any potential morphological effects in the ring structure.

More sophisticated models may be required to investigate these phenomena.  However, since dust produced according to a \citet{d69} crushing law or a \citet{d07} crushing law yields a cloud dominated by the largest grains, as we showed above, we hypothesize that resonant rings in exozodiacal clouds may often be dominated by a single particle size whose PR time is roughly equal to the its collisional time.

\subsection{Ring Detectability}

The detectability of a resonant ring structure depends on many factors specific to the telescope being used and the observing conditions.  We address this complicated issue by imposing one simplifying assumption: a minimum detectable optical depth ring contrast of 1.5.  This assumption likely underestimates the sensitivity of a TPF-like mission to rings in exozodiacal clouds analogous to the solar zodiacal cloud.  In such a cloud, a ring 0.4 AU wide located at 1 AU from the star has $\sim 15$ times the total flux of an Earth-like planet at 1 AU, even for a contrast of unity.  Our assumption, conservative on the basis of photon noise alone, allows for the possibility of unknown systematic noise that could hinder the detection of extended structures.

Figure \ref{minimumMp} shows the minimum detectable planet mass as a function of semi-major axis and maximum dust particle size based on Equations \ref{analyticcontrast2} and a \citet{d69} crushing law.  The masses and semi-major axes of Earth, Mars, and the planet OGLE-2005-BLG-390Lb, detected by the microlensing technique \citep{bbf06}, are marked for reference.  This plot shows that an Earth-mass planet at 1 AU might be detectable if the ring contains grains more than a few tens of microns in size and a planet with mass equal to a few times that of Mars might be detectable near 10 AU if the ring contains grains more than one hundred microns in size.

The detectability of a ring structure depends upon the size distribution of dust within the ring structure.  Dust produced according to a crushing law less steep than the \citet{d69} crushing law ($\alpha < 3.4$) will result in more highly contrasted ring structures because of the increased relative contribution of the large grains.  For crushing laws with $\alpha = 3$ and $\alpha = 2$, the curves of constant maximum particle size shown in Figure \ref{minimumMp} shift downward by a factor of approximately 1.25 and 1.55, respectively.

These values are subject to the assumptions of our simulations, which do not include a dynamically hot component in the dust cloud.  This component would reduce the contrast in the ring, making planets harder to detect for a given cloud mass.  So the detection limits shown in Figure \ref{minimumMp} should be thought of as best-case scenarios.

\section{Caveats}

Our simulations include a number of simplifying assumptions, which we summarize here.  We ignored the effects of dynamically hot dust, like dust that might come from comets.  Trapping probability decreases dramatically for particles on highly eccentric and inclined orbits, so we expect dynamically cold dust to dominate any resonant debris disk structure.  As a first approximation, we can treat the contribution from the dynamically hot dust as a constant surface density cloud component, which reduces the contrast of any structure formed from the dynamically cold component.  Estimates of the ratio of asteroidal dust to cometary dust in our solar system range from 1:10 to 7:10 \citep{i07}.  For other systems, this ratio is also unknown.

Our simulations also assumed a single planet on a circular orbit around a Sun-like star.  We have performed trial simulations of our solar system and demonstrated that the presence of Jupiter may reduce the Earth's ring contrast.  Other multiple-planet systems may also exhibit a similar effect.  Additionally, planets on eccentric orbits give rise to additional MMRs with different capture probabilities and geometries \citep{kh03}.

The ring contrasts of inclined systems can vary significantly depending on the inclination and radial extent of the dust cloud.  In edge-on systems, resonant features can overlap as seen from the Earth, complicating their interpretation.  The contrasts we provide are useful only to systems for which projection effects can be taken into account.

Finally, our multi-particle-size models demonstrate the subtlety of collisional effects in dust clouds.  Collisional effects can determine the relative populations of large and small grains and potentially alter the morphology of the ring structures.  Our simulations can not yet handle these effects in detail.

\section{Conclusions}

We have implemented our own hybrid symplectic integrator for the $n$-body problem and used it to simulate collisionless debris disks, taking into account solar wind and drag effects.  Each simulation contained 5,000 particles.  We found that this number of particles suffices to populate the dominant MMRs of a low-mass planet with an accuracy at the few percent level, yielding for the first time models of the surface brightness distributions of exozodiacal clouds that we can use to quantitatively study the contrasts of resonant features---not just their geometries.

We generated a catalog of resonant structures induced by a single planet on a circular orbit around a Sun-like star, available online at http://asd.gsfc.nasa.gov/Christopher.Stark/catalog.php.  We investigated 120 sets of model parameters, spanning a range of planet masses, planet semi-major axes, and values for $\beta$, assuming dust grains launched from orbits with low $e_{\rm dust}$ and $i_{\rm dust}$.  The resulting ring structures exhibited leading-trailing asymmetries, gaps near the locations of the planets, and sharp inner edges at $\approx 0.83 a_{\rm p}$.

\begin{figure}
\begin{center}
\includegraphics[width=6.5in]{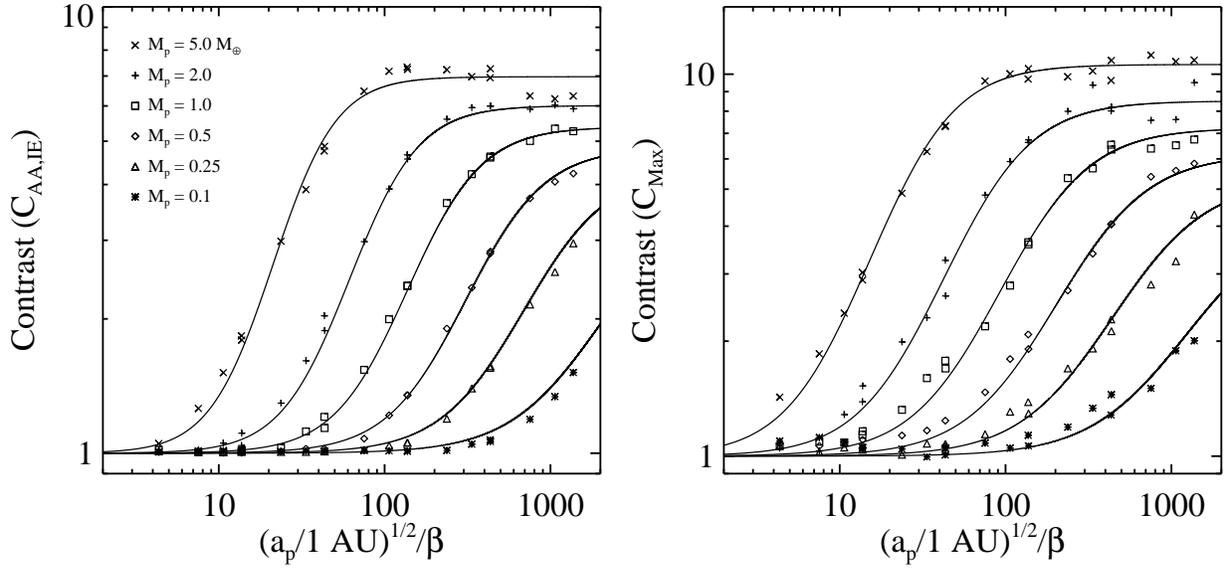}
\caption{The contrast in surface density of the ring structure compared to the background cloud for all combinations of $M_{\rm p}$, $a_{\rm p}$, and $\beta$ (see Section \ref{ringcontrast} for definitions of contrast).  The contrast is only a function of two parameters: planet mass and $\sqrt{a_{\rm p}}/\beta$.  The solid lines are fits to the data (see Equation \ref{contrastequation}).  \label{contrasttrends}}
\end{center}
\end{figure}

We performed a detailed analysis of the surface density contrasts of the rings (Figure \ref{contrasttrends}).  We showed that for a planet on a circular orbit, the contrast and morphology of the rings are to good approximation functions of only two parameters, $M_{\rm p}$ and $\sqrt{a_{\rm p}}/\beta$, for a given stellar mass and distribution of dust sources for $M_{\rm p} \lesssim 5\; M_{\earth}$ and $\beta \lesssim 0.25$.  Equation \ref{contrastequation} summarizes the contrasts of our single-particle-size models as a function of these parameters.  Considering only the dynamically cold particles analogous to particles released by asteroids in the solar system, we find that terrestrial-mass planets are capable of producing resonant ring structures with azimuthally averaged contrasts up to $\sim 7:1$.

By combining our simulations of grains with particular $\beta$ values, we assembled multi-particle-size models of 25,000 particles each.  Releasing the particles according to a \citet{d69} crushing law without any subsequent collisional processing results in composite clouds whose optical depths are dominated by large particles; large particles will dominate images of these clouds in visible light and throughout the IR.  Based on these composite models, we suggested that the best current models for exozodiacal clouds are those with a narrow range of grain sizes corresponding to grains whose collision time roughly equals their PR time.  Future models should account for processes like grain-grain collisions that destroy large grains.

Equations \ref{analyticcontrast} and \ref{analyticcontrast2} provide semi-analytic predictions for the contrast in optical depth of a multi-particle-size cloud of dynamically cold grains.  For ring structures composed of silicate grains released according to a Dohnanyi crushing law ($\alpha = 3.4$), Equation \ref{analyticcontrast2} gives an approximate contrast of
\begin{equation}
	\label{finalcontrastequation}
	\langle C_{\tau{\rm ;AA, IE}} \rangle \approx \left\{ \begin{array}{ll}
		1 + 
		4.4  \left(\frac{M_{\rm p}}{M_\earth}\right)^{0.19} -
		52 \left(\frac{s_{\rm max}}{1\; \mu{\rm m}}\right)^{-0.6}\left(\frac{a_{\rm p}}{1\; {\rm AU}}\right)^{-0.3} \left(\frac{M_{\rm p}}{M_\earth}\right)^{-0.57}
		& \mbox{for $s_{\rm max} > s_2$} \\
		
		\frac{1}{1+\left(\frac{M_{\rm p}}{M_\earth}\right)^{0.18}}
		\left[ 4 \left(\frac{s_{\rm max}}{1\; \mu{\rm m}}\right)^{-0.6}\left(\frac{M_{\rm p}}{M_\earth}\right)^{-0.49} \left(\frac{a_{\rm p}}{1\; {\rm AU}}\right)^{-0.3} \right. & \\
		\indent \left. + \left( 4 \left(\frac{s_{\rm max}}{1\; \mu{\rm m}}\right)^{-0.6} \left(\frac{M_{\rm p}}{M_\earth}\right)^{-0.67}\left(\frac{a_{\rm p}}{1\; {\rm AU}}\right)^{-0.3} \right)^{-\left(\frac{M_{\rm p}}{M_\earth}\right)^{0.18}}\right] & \mbox{for $s_{\rm max} < s_2$} \\
	\end{array} \right.		
\end{equation}
where $\langle C_{\tau {\rm ;AA,IE}} \rangle$ is the ratio of the azimuthally averaged optical depth in the ring structure to the azimuthally averaged background optical depth, $s_{\rm max}$ is the maximum grain size in the ring structure, and $s_2$ is given by Equation \ref{s2equation}.  For the case $s_{\rm max} > s_2$, the first two terms in Equation \ref{finalcontrastequation} represent the contrast in the adiabatic limit.  The remaining term (and the terms in the $s_{\rm max} < s_2$ case) represents deviations from this limit for smaller particles or smaller semi-major axes.

\begin{figure}
\begin{center}
\includegraphics[width=6.5in]{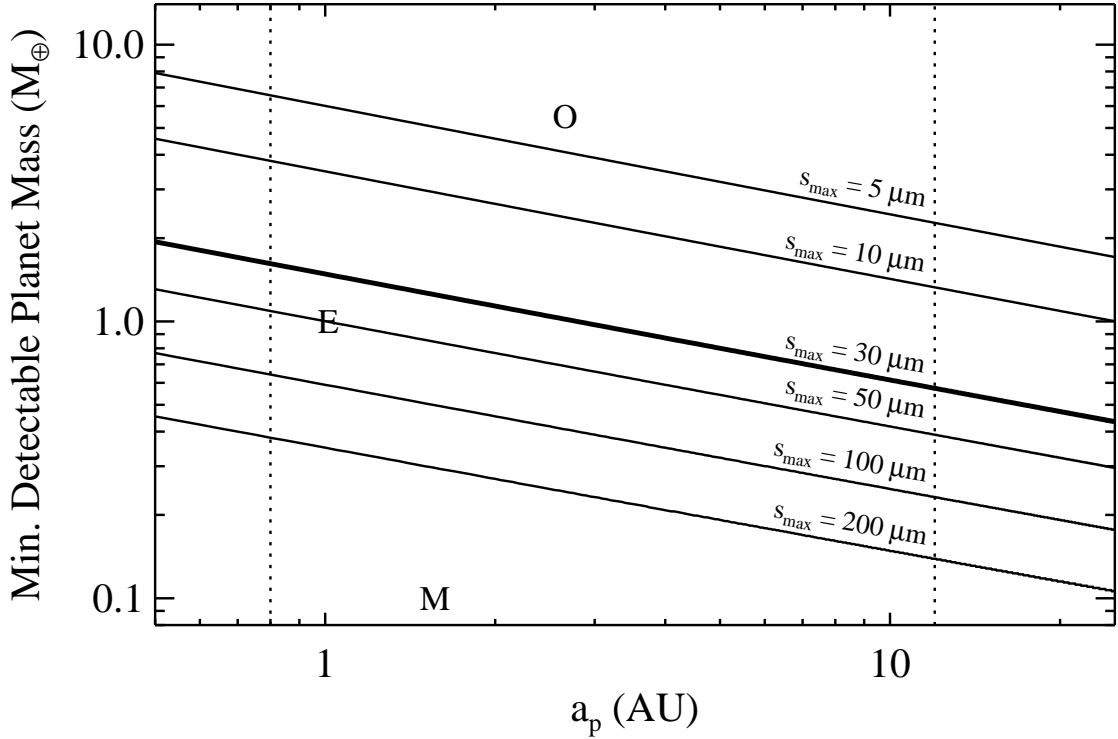}
\caption{Minimum detectable planet mass in a multi-particle-size collisionless cloud as a function of semi-major axis and maximum grain size, assuming a Sun-like star, a minimum detectable ring contrast of $\rm C_{\tau,AA,IE} = 1.5$, and dust produced according to a \citet{d69} crushing law ($\alpha = 3.4$).  Earth-like and Mars-like planets are denoted with an $E$ and $M$, respectively.  The $5.5\, {\rm M}_{\earth}$ exoplanet OGLE-2005-BLG-390Lb is denoted with an $O$ \citep{bbf06}.  Listed values for maximum dust size in the ring structure assume perfectly absorbing spherical grains with mean density $\rho = 2.0$ gm cm$^{-3}$ and radius $s_{\rm max}$.  The bold line shows the case of the solar zodiacal cloud, for which the observed emission is dominated by 30 $\mu\rm{m}$ grains \citep{fd02}.  The dashed lines show typical inner and outer detection limits for a mission similar to TPF.\label{minimumMp}}
\end{center}
\end{figure}

We plotted the mass of the smallest planet that could be detected through observation of a resonant ring structure as a function of planet semi-major axis and particle size in Figure \ref{minimumMp}.  We assumed a cloud composed of a range of particle sizes adhering to a \citet{d69} crushing law and a minimum detectable optical depth contrast of 1.5:1.  We found that planets with masses just a fraction of the Earth's may form detectable ring structures if the rings harbor grains more than several tens of microns in size.

\acknowledgments

We would like to thank the National Aeronautics and Space Administration and Goddard Space Flight Center for support of this research through funding from the Graduate Student Researchers Program, and the Harvard-Smithsonian Center for Astrophysics and NASA's Navigator Program for their financial support via the Keck Interferometer Nuller Shared Risk Science Program.


\newpage
\clearpage

\newpage
\clearpage

\newpage
\clearpage

\newpage
\clearpage

\newpage
\clearpage

\newpage
\clearpage

\newpage
\clearpage

\newpage
\clearpage

\newpage
\clearpage

\newpage
\clearpage

\newpage
\clearpage

\newpage
\clearpage

\end{document}